\documentstyle[12pt]{article}
\begin{document}
\baselineskip 18pt
\def\today{\ifcase\month\or
 January\or February\or March\or April\or May\or June\or
 July\or August\or September\or October\or November\or December\fi
 \space\number\day, \number\year}
\def\thebibliography#1{\section*{References\markboth
 {References}{References}}\list
 {[\arabic{enumi}]}{\settowidth\labelwidth{[#1]}
 \leftmargin\labelwidth
 \advance\leftmargin\labelsep
 \usecounter{enumi}}
 \def\newblock{\hskip .11em plus .33em minus .07em}
 \sloppy
 \sfcode`\.=1000\relax}
\let\endthebibliography=\endlist
%
%
\begin{titlepage}
\begin{flushright} 
           ICRR-Report-391-97-14 \\
           OCHA-PP-97 \\ 
           kure-pp/97-01 \\
\end{flushright} 
\  \
\renewcommand{\thefootnote}
{\fnsymbol{footnote}}
\begin{center}
{\large {\bf Study of $CP$ Violation: }}  \\
{\large {\bf Electroweak Baryogenesis and Anomalous $W$-Boson Couplings 
\footnote{To appear in the Proceedings of the KEK
meetings on '$CP$ violation and its origin' ($1993-1997$).
}
}} 
\vskip 0.5 true cm
\vspace{1cm}
Akio Sugamoto $^1$, Mayumi Aoki $^1$  
\footnote{Research Fellow of the Japan Society 
for the Promotion of Science.}, 
Tomoko Kadoyoshi $^1$ , Miho Marui $^2$,  \\
Noriyuki Oshimo $^3$,  
Tomomi Saito $^1$, Tomoko Uesugi $^1$, Azusa Yamaguchi $^4$ \\ 
\smallskip
(Ochanomizu $CP$ Study Group)
\\
\vskip 0.5 true cm 
{\it $^1$Department of Physics {\rm and} 
 Graduate School of Humanities and Sciences,   
Ochanomizu University, Tokyo 112, Japan \\ 
 $^2$Faculty of Social Information Science, Kure University, 
                  Hiroshima 724-07, Japan \\
     $^3$Institute for Cosmic Ray Research, 
            University of Tokyo, Tokyo 188, Japan  \\
     $^4$The Institute of Physical and Chemical Research, 
               Saitama 351-01, Japan 
 }
\end{center}

\vskip 0.5 true cm

\centerline{\bf Abstract}
\medskip
    The contributions from the Ochanomizu $CP$ Study Group to
the KEK meetings on "$CP$ violation and its origin" ($1993-1997$) 
are summarized on electroweak baryogenesis and anomalous 
$W$-boson couplings.  
We survey planned new experiments which could examine some aspects 
studied in our contributions.   
We also discuss several issues on baryogenesis.   
Ten problems are presented for further studies.  

\vskip 0.5 true cm

\end{titlepage}
\newpage 

\section{Introduction}

     From February 1993 to March 1997, we had five meetings at KEK on "$CP$ 
violation and its origin" once a year. This series of meetings was
especially fruitful to our group:
We presented the works so far done, and got, from discussions
and talks, critical or helpful advices and in particular new
ideas for our studies the next year and
the year after.

     The talks of our group given at the meetings were as follows:  

$\bullet$ The first meeting ('93) \\
1) Azusa Yamaguchi: Electroweek baryogenesis and the phase transition
dynamics \cite{azusa}.

$\bullet$ The 2nd meeting ('94) \\
2) Akio Sugamoto: Electroweak phase transition and the
baryogenesis \cite{azusa,akio}.

$\bullet$ The 3rd meeting ('95) \\
3) Akio Sugamoto: A memory of Yoshiki Kizukuri \cite{akio2}.\\ 
4) Tomomi Saito: $CP$-odd form factors of the $WWZ$ coupling with a singlet
quark \cite{tomomi}.

$\bullet$ The 4th meeting ('96) \\
5) Noriyuki Oshimo: The electric dipole moments of the $W$ boson 
and the neutron in the supersymmetric model \cite{noriyuki}. \\ 
6) Mohammad Ahmady: Rare $B$ decays \cite{ahmady}. \\ 
7) Tomoko Uesugi: Baryogenesis in the vector-like quark model \cite{tomoko}. 

$\bullet$ The last meeting ('97) \\
8) Emi Kou: Combined $B \rightarrow X_s \psi$ and $B \rightarrow X_s
\eta_c$ as a test of factorization \cite{emi}. \\ 
9) Mayumi Aoki: Electroweak baryogenesis from chargino transport 
in the supersymmetric model \cite{mayumi}.

     Among the works presented in these talks, we summarize 
those on electroweak baryogenesis and anomalous $W$-boson couplings, 
together with their related works by us.  
A summary of the talks 6) and 8) is given separately by
Ahmady and Kou \cite{ahmady2}. 
Works on the electric dipole moments (EDMs) of the neutron and 
the electron presented in the talks 5) and 9) and in the talks by 
Kizukuri \cite{kizukuri} are also 
separately summarized \cite{mayumi2}.  

     Some aspects discussed in our contributions could be examined 
in planned new experiments.  From this point of view, we survey 
experiments for the neutron EDM using ultracold neutrons, 
experiments for $CP$ asymmetries at $B$ factory, and 
long baseline experiments for neutrino oscillations.    
We also take this occasion to discuss several issues on 
electroweak baryogenesis which we think are important 
for further studies.   

\section{Electroweak Baryogenesis}

     We have studied baryogenesis so far in three models which are extended 
from the standard model (SM). Here we
summarize our works in terms of Sakharov's three criterions, namely, 1) $CP$
violation, 2) thermal non-equilibrium, and 3) baryon number violation,
accompanied with 4) the results and 5) the characteristic points.  

\subsection{Model with heavy neutrinos [1,2]}

     1) Right-handed neutrinos $N_i$ and a Majoron field $\phi$ 
are introduced.  The origin of $CP$ violation is   
the complex mass matrix for the neutrinos 
having both Dirac and Majorana masses: 
\begin{equation}
\left(\matrix{\overline{\nu^c_1} & \overline{\nu^c_2} & \ldots & 
                             \overline{N_1} & \overline{N_2} & \ldots}
\right) 
\left(\matrix{ 0 & \lambda_D v(x) \cr
                      \lambda^T_D v(x) & \lambda_M v(x) }
\right) 
\left(\matrix{\nu_1 \cr
                       \nu_2 \cr
                       \vdots \cr
                        N^c_1 \cr
                        N^c_2 \cr
                       \vdots}
\right), 
\label{neutrino mass matrix}
\end{equation}
where we have assumed that the vacuum expectation value of 
$\phi$ and that of the ordinary Higgs field have the same 
position-dependent shape $v(x)$ in the bubble wall.    \\
2) Thermal non-equilibrium is accomplished in the  
first order phase transition with the nucleation and 
expansion of bubbles, for which the order parameter is $\phi$.  
The bubble wall is an interface of the broken phase and 
the unbroken phase.  \\
3) Lepton number $N_L$ is not conserved inside the bubble wall, which 
induces the reflection from the wall $\nu_i\rightarrow \bar\nu_j$ 
and its $CP$ conjugate process $\bar\nu_i \rightarrow\nu_j$. 
Owing to $CP$ violation, 
these processes have probabilities different from each ohter.  
The produced non-vanishing value for $N_L$  
subsequently induces a non-vanishing value for baryon number $N_B$ 
through fast $N_B$-violating transitions generated 
by electroweak anomaly.  \\ 
4) For thermally generated neutrino masses of $50-100$ GeV 
with the phase transition temperature of 100 GeV,  
the ratio of baryon number to entropy 
is consistent with its observed value, if $CP$ violation 
is $10^{-5}-10^{-7}$ measured by the Jarskog parameter $J$ for the
neutrino masses.  \\
5) In the unbroken phase, 
the Majorana mass terms for neutrinos vanish  
and $N_B-N_L$ are conserved. Thus the net density 
of $N_L$ leads to the net density of $N_B$. 
The temporal development of the phase transition was simulated, 
which shows that the temporal
change of the expansion velocity $v_w$ and the fusion effect of 
nucleated bubbles are important.  
Depending them, the final result could be modified considerably.

\subsection{Model with vector-like quarks [7]}

     1) A vector-like quark $U$ with the charge 2/3 
and a Higgs singlet $S$ are introduced.   
The vacuum expectation value $V(x)$  
of $S$ has a position-dependent complex phase in the bubble wall,  
which gives a complex mass matrix for the up-type quarks: 
\begin{equation}
\left(\matrix{\ldots & \overline{t_R} & \overline{U_R}}
\right) 
\left(\matrix{\ddots & & \cr
                          & fv(x) & 0 \cr
                  & F V(x) + F' V^*(x) & \mu} 
\right)
\left(\matrix{\vdots \cr
                         t_L \cr
                         U_L \cr }
\right). 
\label{vector-like quark mass matrix}
\end{equation}
The complex value for $V(x)$ could be induced spontaneously 
from the $CP$-conserving Higgs potential with a Higgs doublet and $S$. \\ 
2) During the electroweak phase transition the bubble formation is 
expected, for which the order parameter is the ordinary Higgs field.  
The first order phase transition is the origin of 
thermal non-equilibrium. \\
3) Difference of the reflection and transmission probabilities between
the processes $t, U \rightarrow t, U$ 
and their $CP$ conjugate processes  
$\bar{t}, \bar{U}\rightarrow \bar{t}, \bar{U}$ at the wall 
produces a net flux of hypercharge outside the bubble.  
Then a non-vanishing value of the chemical potential $\mu_{B}$ 
for baryon number, 
i.e. "thermal pressure" to increase baryon number, is generated,  
which triggers the $N_B$-violating transitions to produce baryon number.  \\ 
4) For the $U$-quark mass of $300-500$ GeV with the wall
velocity $v_w= 0.1-0.5$ and the wall width $\delta_w$=(100 GeV)$^{-1}$ 
the baryon number to entropy ratio of $10^{-10}-10^{-11}$ 
can be reproduced.  \\
5) The model is economical as an extension of the SM, and the
mechanism of spontaneous $CP$ violation can be embedded.  
Spontaneous $CP$ violation, however, should be adopted with care 
because of the appearance
of both $CP$-even and $CP$-odd bubbles. In order to keep one type 
of bubble and eliminate another, tiny but explicit $CP$ breaking
of order $10^{-16}$ is necessary.
 
\medskip
{\bf [Problem 1]} 
     Study how to measure directly various types of $CP$ violation
originating from the Higgs potential. Is it possible to distinguish      
experimentally spontaneous $CP$ breaking mechanism from explicit one? 

\subsection{Supersymmetric standard model [9,13]}

     1) An extra $CP$-violating phase is  
introduced in the mass matrices of gauginos $\lambda$ 
and Higgsinos $\psi$, the spin 1/2
superpartners of gauge and Higgs bosons, respectively. For the charged
sector the mass matrix is given by 
\begin{equation}
\left(\matrix{\overline{\lambda^-} & \overline{(i\psi^+_2)^c}}
\right)
\left(\matrix{\tilde m_2 & -gv_1^*(x)/\sqrt{2} \cr
                -gv_2^*(x)/\sqrt{2} & m_H}        
\right)    
\left(\matrix{\lambda^- \cr
                       i\psi_1^-}
\right),
\label{chagino mass matrix}
\end{equation}
where $v_1(x)$ and $v_2(x)$ denote the vacuum expectation values 
for the two Higgs doublets;  
$\tilde m_2$ the soft supersymmetry-breaking mass for the SU(2) gauginos; 
and $m_H$ the Higgsino mass parameter in  
the mixing term of the two Higgs doublet superfields.  
Among the four parameters in the mass matrix, one complex
phase $\theta$ may remain in $m_H$, escaping from the rephasing of 
field variables. This is the origin of $CP$ violation intrinsic in  
the supersymmetric standard model (SSM) useful for baryogenesis.   \\
2) The first order phase transition with the bubble formation is
the origin of thermal non-equilibrium. \\ 
3) Difference of the probability between $CP$ conjugate processes for 
the reflections and transmissions 
of gauginos and Higgsinos at the bubble wall 
generates net hypercharge in the
unbroken phase outside the bubble. The produced hypercharge 
density makes the chemical potential $\mu_{B}$ non-vanishing, 
which triggers the $N_B$-violating transitions for baryogenesis.  \\
4) Having the new $CP$-violating phase not much small, $\theta >0.1$, the
observed ratio of baryon number to entropy can be excellently reproduced, 
if the mass parameters $\tilde m_2$ and $|m_H|$ are of order 100 GeV.   
The wall velocity and the wall width have been taken 
respectively for $v_w= 0.1-0.6$ 
and $\delta_w= 1/T- 5/T$, $T$ being the phase transition temperature.   \\ 
5) The most interesting point of this scenario is the following:  
The value of $CP$-violating phase explaining the baryon number to 
entropy ratio of our universe is located in the range which is attainable 
in near future experiments
for the EDM of the neutron ($D_n$). 
The value of $D_n$ is predicted to be $10^{-25}-10^{-26} e\mbox{cm}$.   
Top squarks, instead of gauginos and 
Higgsinos, could also 
assume the same role for baryogenesis.  

\section{Anomalous $W$-boson Couplings}

     Anomalous $W$-boson couplings for the $WWZ$ and 
$WW\gamma$ interactions have been studied 
in the models discussed 
also for baryogenesis in the preceding chapter.     
The $WWZ$ or $WW\gamma $ vertex can generally be expressed as 
\begin{eqnarray}
\Gamma^{\nu\lambda\mu}(q, \bar{q}, P) &=& 
       f_{1} (q-\bar{q})^{\mu} g^{\nu\lambda} 
     - f_{2} (q-\bar{q})^{\mu} P^{\nu} P^{\lambda}/m_{W}^{2} 
     + f_{3} ( P^{\nu} g^{\mu\lambda} - P^{\lambda} g^{\mu\nu} ) 
                      \nonumber \\ 
 & & + i f_{4} (P^{\nu} g^{\mu\lambda} + P^{\lambda} g^{\mu\nu} ) 
     + i f_{5} \varepsilon^{\mu\nu\lambda\rho} (q - \bar{q} )_{\rho} 
     - f_{6} \varepsilon^{\mu\nu\lambda\rho} P_{\rho}  
                      \nonumber \\
 & & - f_{7} ( q - \bar{q} )^{\mu}\varepsilon^{\nu\lambda\rho\sigma}
                 P_{\rho} (q-\bar{q} )_{\sigma}/m_{W}^{2}. 
\end{eqnarray}
Among the seven form factors, $f_1$, $f_2$, $f_3$, and $f_5$ 
are $CP$-even and $f_4$, $f_6$, and $f_7$ are $CP$-odd.   
In the SM the form factors $f_1$ and $f_3$ alone do not vanish 
at the tree level, which holds also in the three models we discuss.  
Some of these form factors will be investigated through the process 
$e^+e^- \rightarrow W^+W^-$ at LEP2 or planned linear colliders.  
In particular, since the $CP$-odd form factors vanish 
at both the tree level and the one-loop level in the SM,   
new physics could be examined unambiguously through their effects.   

\subsection{Model with heavy neutrinos [14]}

     Radiative corrections to the $CP$-even form factors 
coming from the one-loop 
diagrams with the heavy neutrinos have been studied.   
Around the threshold for the pair production of neutrinos 
the behaviors of these form factors depend on whether 
the type of the neutrinos is Dirac or Majorana.   
The difference of cross section for 
the process $e^+e^- \rightarrow W^+W^-$ could be of order 
$10^{-4}$ between the process involving Dirac neutrinos 
and that involving Majorana neutrinos 
for their masses of $100-1000$ GeV.   
It will be possible to examine the parameters in the
neutrino mass matrix in Eq. (\ref{neutrino mass matrix}). 

\subsection{Model with vector-like quarks [4]}

     A vector-like quark $D$ with the charge $-1/3$ is introduced,     
and $CP$ invariance is violated either explicitly or spontaneously.  
The mass matrix for the down-type quarks is given by an 
equation similar to Eq. (\ref{vector-like quark mass matrix}), 
which induces both $CP$ violation and 
flavor changing neutral current (FCNC).  
The SU(2) neutral current $J^3_\mu$ for the down-type 
quarks is given by 
\begin{equation}
J^3_\mu = - \frac{1}{2}\sum^3_{\alpha, \beta = 1}
\left(\delta_{\alpha\beta} - V_{4\alpha}^* V_{4\beta} \right)
\bar{d}^\alpha_L \gamma_\mu d^\beta_L, 
\label{FCNC}
\end{equation}
where $V$ denotes the unitary matrix which diagonalizes the mass 
matrix of the down-type quarks.    
This current gives non-vanishing values to $f_4$ and $f_6$ for 
the $WWZ$ vertex 
at the one-loop level, while $f_7$ remains vanishing.  
It turned out, however, that the induced form 
factors are both much smaller than $10^{-4}$ for the $D$-quark mass of 1 TeV, 
which would be difficult to be detected in the near future.  

\subsection{Supersymmetric standard model [5,15]}

     The interactions for the chargino, neutralino, and $W$ boson 
violate $CP$ invariance, which induce the $CP$-odd form factors 
at the one-loop level.  For the $WW\gamma$ vertex the magnitude 
of $f_6$ becomes of order $10^{-4}$, while $f_4$ and $f_7$ 
vanish.  Although the induced form factor is not so large, 
it could be attainable experimentally in the near future.
The $WWZ$ vertex has also been studied, 
obtaining the results $f_4, f_6\sim 10^{-4}-10^{-6}$, $f_7=0$.   

\medskip
{\bf [Problem 2]} Study various methods to detect small $CP$
violation effects in the process $e^{+}e^{-} \rightarrow W^{+}W^{-}$.

\section{Prospects for New Experiments} 

     For any theory in physics, experimental verification is 
indispensable.   
Even though a theoretical idea is beautiful
and profound, it is not regarded as true before its experimental evidence
comes out.  We outline several interesting new 
experiments relevant to our contributions, hoping that 
some of the predictions could be examined.    

\subsection{Experiments of neutron EDM}

     In the SSM the baryon asymmetry 
of our universe implies that the EDM of the neutron is not 
much smaller than its present experimental upper bound.  
The improvement of its experimental measurement by one order of magnitude  
could provide an important test for the baryogenesis by the SSM.  
The neutron EDM gives constraints also on the 
model with vector-like quarks \cite{katsuji}.   

     The technique of measuring the neutron 
EDM is progressing very rapidly.  
It was fortunate that this series of meetings was organized as 
a part of the big project on "the ultracold neutron".  
We learned various aspects for the neutron EDM 
experiments \cite{masaike and shimizu}. 
The experiment is now performed at the Institut
Laue-Langevin in Grenoble:  The cold neutron is guided by 
the Ni tube (totally reflecting "fiber for neutron") 20 m upwards, 
reducing the kinetic energy gravitationally.  
Then the neutron is scattered by the recessively rotating 690
(totally reflecting) Ni blades of the so-called "turbines", being further
cooled down to the ultracold neutron (UCN) \cite{ILL1}. 
Let us remind ourselves of the high
school physics that a non-relativistic particle with velocity $V$ stops,
after being scattered by a perfectly reflecting wall which recedes with
velocity $V/2$.  The electric field $E$ of 10 kV/cm is applied to the
obtained UCNs, in parallel or anti-parallel with the magnetic field $B$
of 1 $\mu$T, measuring the change of the magnetic resonance frequency: 
\begin{equation}
\nu \pm \Delta \nu = - 2(\mu_n\cdot B\pm D_n\cdot E)/h 
  \simeq 30~\mbox{Hz} \pm \Delta\nu .  
\end{equation}
The upper bound on $\Delta\nu$ has been obtained as 
$\Delta\nu < 1 ~\mu\mbox{Hz}$ \cite{ILL2}, 
which gives the bound on the neutron EDM $|D_n|< 10^{-25} e$cm.  
This result is consistent with the measurement at 
the Leningrad Nuclear Physics Institute \cite{LNPI}.   

     The improvement of the measurement could be done, if a 
more effective source of UCNs is available.  A hopeful candidate 
is the superthermal method \cite{yoshiki}:  
The cold neutron with wavelength about 9 {\AA} is guided into the liquid 
$^4$He at 1 K.  Then the neutron stops, 
by spending all its kinetic energy and momentum
on a phonon emission, since the disperson curves of neutron and phonon
coincide at the wavelength.   
We hope that this technique will be successfully used in the 
neutron EDM measurement in the near future.
The improvement is also possible by a more
accurate monitoring system of the magnetic field which tends to fluctuate
randomly and thus becomes the origin of a large systematic error. 

\medskip
{\bf [Problem 3]} If there is a possibility of measuring $D_{n}$
very precisely to the order of $10^{-30}e$cm by e.g. some neutron
interferometer, how much can one say about models beyond the standard
model?

\subsection{$B$ factory experiments}

     The model with vector-like quarks  
has FCNC interactions at the tree level as seen in Eq. (\ref{FCNC}) 
as well as the breaking of the unitarity triangles,  
which could be observed in experiments at $B$ factories.  
In these experiments the time dependent $CP$ asymmetry 
\begin{equation}
A(t) = \frac{\Gamma(B^0\rightarrow f) - \Gamma(\bar{B^0}\rightarrow f)}
{\Gamma(B^0\rightarrow f) + \Gamma(\bar{B^0}\rightarrow f)} 
\simeq\pm\sin(\Delta M_B t) \sin2(\phi_M+\phi_D)
\label{asymmetry}
\end{equation}
can be measured for $B$-meson decay.  The results   
may differ drastically from the SM predictions  
in various modes, such as $B_d\rightarrow\psi K_S, D^+D^-, \pi^+\pi^-$,  
and $B_s\rightarrow\rho K_S$ \cite{morozumi}.   
In the SSM, although effects of the new sources of $CP$ violation  
are small, new contributions to $B$-$\bar B$ mixing indirectly 
affect the $CP$ asymmetries \cite{noriyuki2}.  

     The $B$ factory experiments will begin  
in the near future at KEK and SLAC.  The KEK $B$ factory 
will operate the asymmetric $e^+e^-$ collider of the electron 
energy 8 GeV and the positron energy 3.5 GeV with luminosity 
${\cal L}=2\times10^{33} \mbox{cm}^{-2}\mbox{s}^{-1}$. 

\subsection{Neutrino oscillation experiments} 

     The introduction of masses for the neutrinos leads to 
neutrino oscillations as well as $CP$ violation.   
A test of $CP$ violation in the neutrino sector 
could be performed through neutrino oscillations   
in long baseline experiments \cite{arafune}, by    
measuring e.g. the difference between the transition 
probabilities  
\begin{equation}
A(\nu) = P(\nu_\mu\rightarrow\nu_e) - P(\bar\nu_\mu\rightarrow\bar\nu_e). 
\label{neutrino asymmetry}
\end{equation}
In order to have detectable neutrino oscillations, 
two neutrinos having different
momenta should interfere effectively in the distance $L$.  
The phase difference is therefore required to satisfy the condition  
\begin{equation}
\Delta \phi(L) = |p_1 - p_2|L \simeq \frac{L}{2E}|m_1^2 - m_2^2|\sim 1,
\end{equation}
which is really the case for $L\simeq 100$ km, $E\simeq 1$ GeV, and 
$|m_1^2-m_2^2|\simeq 10^{-2} ~{\mbox eV}^2$.
Then the magnitude of $A(\nu)$ could be of order $10^{-2}$ 
for the parameter values explaining the solar and the atmospheric 
neutrino anomalies. 

     The long baseline experiment is planned at KEK-Kamiokande, 
where neutrinos are produced by the upgraded PS at KEK and detected 
by the water Cherenkov counter at Kamiokande 250 km apart. 
The neutrinos coming from
the proton machine are mainly $\nu_\mu$'s and $\bar\nu_\mu$'s 
produced by $\pi$-mesons or $K$-mesons, so that the asymmetry 
$A(\nu)$ of Eq. (\ref{neutrino asymmetry}) could be measured.   
There is however one problem.  
The neutrino oscillations suffer the matter effects by the  
electrons in the earth, so that the results do not 
only reflect vacuum oscillations but also experimental circumstances.   
Since the latter effects contribute with different signs to the
$\nu$ and the $\bar{\nu}$ oscillations, $A(\nu)$ contains twice as
much the matter effects. 
For subtracting them, the difference of energy dependence between  
the vacuum oscillations and the matter effects may be useful.  

\medskip
{\bf [Problem 4]} Invent a method to detect $CP$ violation
related to heavy neutrinos of order 100 GeV in high energy experiments. 

\section{Discussions}

     The scenario of electroweak baryogenesis contains various 
issues for which further analyses are indispensable to obtain 
accurate results.  
We discuss some of them in an elementary and naive way, 
which we believe is useful for our studies.

\subsection{Baryon number violation}

     In any model of electroweak baryogenesis, the
main source of baryon number violation is the chiral gauge anomaly 
\begin{eqnarray}
\partial_{\mu} J^{\mu}_{B} &=& \partial_{\mu} J^{\mu}_{L} = - N_{g}
\frac{1}{32 \pi^{2}} F^{a}_{\mu\nu}\tilde{F}^{a\mu\nu} = - N_{g}
\partial_{\mu} K^{\mu},  \\
K^{\mu} &=& \frac{1}{32 \pi^{2}} \epsilon^{\mu\nu\lambda\rho}
(F^{a}_{\nu\lambda}A^{a}_{\rho} - \frac{2}{3} \epsilon_{abc}
A^{a}_{\nu}A^{b}_{\lambda}A^{c}_{\rho}),  \nonumber 
\label{anomaly}
\end{eqnarray}
where $N_{g}$ represents the number of generations and 
$K^\mu$ is the Chern-Simons current.   
By this anomaly, baryon and lepton numbers can change while 
$N_B-N_g N_{CS}$ and $N_L-N_g N_{CS}$ being conserved, 
where $N_{CS}$ denotes the Chern-Simons number given by 
\begin{equation}
 N_{CS} = \int {\rm d}^3{\bf x}K^0. 
\end{equation}
Various configurations of the gauge fields give various values for
$N_{CS}$, with different energies 
\begin{equation}
E = \frac{1}{g^2}\int {\rm d}^3{\bf x} \{(F^a_{0i})^2 + (F^a_{ij})^2 \}. 
\end{equation}
Since $N_{CS}$ becomes some integer for $E=0$, 
the vacuum can be classified by $N_{CS}$ as 
$|\mbox{vac}, N_{CS}>$. 

     The baryon and lepton numbers shift through some  
configurations:    
$N_B\rightarrow N_B+N_g$, $N_L\rightarrow N_L+N_g$ with 
$|\mbox{vac}, N_{CS}>\rightarrow |\mbox{vac}, N_{CS}+1>$.   
One of such configurations is called "sphaleron" \cite{manton}, 
which has $N_{CS}=1/2$ and $E=4\pi v/g^2\simeq 10$ TeV.  
There are also a lot of other configurations to generate the changes of 
$N_B$ and $N_L$.  
Nambu's electroweak string is one example \cite{nambu}. 
These are, however, solutions in
the broken phase after the electroweak phase transition ends 
and the transition rates between different baryon numbers 
are negligibly small for baryogenesis.  

     Baryogenesis is undertook not after the phase transition,
but in the course of the phase transition, since baryon number 
violation is not suppressed in the unbroken phase.   
The transition rate  between different baryon numbers has been estimated 
by computer simulations in lattice gauge theory:  
Observing the temporal change of $N_{CS}(t)$, 
the "lifetime" of having a nearly constant value (plateau) for 
$N_{CS}(t)$  is equated 
with $(\Gamma_{N_B-violation}V)^{-1}$ \cite{ambjorn}, 
$V$ being the space volume.  
The obtained transition rate $\Gamma_{N_B-violation}$ per unit 
volume and unit time is given by 
\begin{equation}
\Gamma_{N_B-violation}= \kappa (\alpha_wT)^4, 
\label{sphaleron rate}
\end{equation}
with $\kappa=0.1-1$.   The reason why the rate is a function 
of the product $\alpha_{w}T$ 
is easily understood.  The related Boltzmann weight of the
pure gauge system is a function of $\alpha_w T$, read from 
$\exp(-E/T) = \exp (-\mbox{constant}/g^2T)$. Therefore, the rate per unit
time and unit volume may be written as Eq. (\ref{sphaleron rate}) 
from the dimensional analysis.  

     The $N_B$-violating transition rate implies that one
transition occurs typically in the domain with the volume 
$(\alpha_{w} T)^{-3}$ and the time interval $(\alpha_{w} T)^{-1}$:  
Each domain has 
a configuration of the gauge fields with a value of $N_{CS}$.  
Each configuration appears with a proper Boltzmann weight. 
This is something like the bubble nucleations, although more
complicate, since there are infinite types of bubble (infinite integers
for the vacua and infinite real numbers for the excited configurations).
If the quarks or leptons enter this medium, what really happens
afterwards?

\medskip
{\bf [Problem 5]} What is the physical picture of the baryon number
generating mechanism in the unbroken phase at finite temperature? 

\subsection{Phase transition}

     In the electroweak baryogenesis, 
thermal non-equilibrium is attributed
to the first order phase transition of the universe accompanied by  
the electroweak symmetry breaking. 
Bubbles of the broken phase nucleate, expand, fuse with
each other, and finally cover the whole space. 
Static properties of the phase transition can be obtained 
by analyzing the effective
potential for the classical Higgs fields at
finite temperature at least in the one-loop approximation. 
In the SM the one-loop effective potential is written 
in the large temperature approximation by 
\begin{equation}
V_{T}(\phi) = \frac{\lambda_{T}}{4} \phi^{4} - E T \phi^{3} + D ( T^{2} -
T^{2}_{0} ) \phi^{2},
\end{equation}
where $T$ denotes temperature.  
The first order phase transition could be achieved, 
if the potential has the $\phi^3$ term.  
A bump then exists in the effective potential and separates 
the unbroken and the broken minima. 
The global minimum of the potential is at $\phi=0$ for $T>T_c$ 
and at $\phi=\phi_{min}\neq 0$ for $T<T_c$, where the 
critical temperature $T_c$ is given by 
\begin{equation}
T_{c}= \frac{T_0}{\sqrt{1 - E^2/\lambda_{T_c}D}}.  
\end{equation}
The $\phi^3$ term is induced exclusively by the interactions of bosons 
and the order parameter $\phi$~\cite{dolan}.
Therefore, the introduction of additional Higgs fields could  
increase the coefficient $E$,
which raises the height of the bump and makes the first order phase
transition stronger. 
This is one reason why multi-Higgs models are adopted 
in various attempts for baryogenesis. 
However, it necessitates elaborate analyses to find in various models 
whether the electroweak phase transition is of the 
strongly first order or not. 

\medskip
{\bf [Problem 6]} In multi-Higgs models, various
kinds of bubbles may coexist having different vacuum expectation 
values for the Higgs fields. Discuss the complexity of the  
phase transition.  

\medskip
     As the temperature $T$ cools down from the critical 
temperature $T_c$, the broken phase
becomes more energetically favorable than the unbroken phase. 
When the temperature becomes lower than $T_{c}$ for a certain amount 
and the latent heat $\epsilon$ becomes nonnegligible, 
the phase transition begins by nucleations of bubbles with 
the broken phase.  
The latent heat is the difference of the effective
potential between the broken and unbroken minima, 
being released during the transition. 
The shapes of the nucleated bubbles are determined by  
stationary configurations for the free energy $F$ of the 
Higgs fields roughly given by 
\begin{equation}
F= \int{\rm d}^3{\bf x}[\frac{1}{2} Z_{T} (\phi) (\nabla \phi)^{2} 
              + V_{T} (\phi)].  
\end{equation}
The free energy of the bubble is essentially the sum of 
the surface energy $4\pi R^2\sigma$ and   
the energy $-(4/3)\pi R^3\epsilon$ inside the bubble coming from
the latent heat.  
Approximating that the width of the bubble wall 
is much smaller than the radius of the bubble, 
the width $\delta_{w}$ and the radius $R_{c}$ 
of the critical bubble are given by  
\begin{eqnarray}
\delta_{w} &\approx& \frac{\sqrt{\lambda_{T}}}{ET}, 
\quad R_{c}\approx \frac{(ET)^{3}}{\sqrt{\lambda_{T}^{5}} \epsilon}, \\
 \epsilon &\approx& D (T_{c}^{2} - T^{2} ) \phi_{min}^{2}.   \nonumber 
\end{eqnarray}
The critical bubble is the maximum point for the free energy, 
which is something like the sphaleron solution, 
connecting the smaller shrinking bubble and
the larger expanding bubble.
A problem is that we have essentially only one dimensional quantity,
namely, the temperature $T$. Ignoring the details of the model, the wall width
and the critical radius are of order $10^{-18}$ m.  
To have a clear interface of the bubble, 
i.e. $\delta_{w} \ll R_{c}$, the model has to give a large value 
to $E$ or a small value to $\epsilon$.   

     The nucleation rate $I$ of the critical bubble per unit volume and unit
time is given in terms of the free energy $F_{c}$ of the critical bubble as
\begin{eqnarray}
I &=& T^{4} \left( \frac{F_{c}}{2\pi T} \right)^{3/2} \exp \left(-
\frac{F_{c}}{T} \right) ,  
\label{nucleation rate} \\
\frac{F_{c}}{T} &\propto& \frac{1}{ (T_{c}^{2}/T^{2} -1)^{2} }.
\nonumber 
\end{eqnarray}
This equation shows that the nucleation rate 
is almost vanishing during some interval of cooling down from $T_{c}$, 
but rapidly increases to become nonnegligible at some temperature
$T=T_c-\Delta$.  
The phase transition occurs when the nucleation rate becomes 
larger than the expansion rate of the universe, i.e. $F_c/T<140$.   
However, how can the exact value of $\Delta$ be determined?  

     The phase transition, if it is weakly first order, 
may proceed even at the temperature above $Tc$
through so-called "subcritical bubbles".     
The contributions of the subcritical bubbles to the phase 
transition have been studied, including both quantum and 
thermal fluctuations at finite temperature \cite{tomoko2}. 
If the subcritical bubbles 
live long, the phase transition dynamics could drastically 
be changed from its conventional picture.   
However, a recent study suggests that the subcritical bubbles are easily
deformed from spherical shape and 
non-spherical bubbles are generally short-lived \cite{chinami}. 

     The electroweak phase transition has also been studied 
by computer simulation in lattice gauge theory.   
In a recent work \cite{aoki}, where the Monte Carlo time history
of the Higgs action $L_{s}(t)$ coupled to the gauge fields is simulated,
we can see that appearances and disappearances of bubbles do occur 
in the course of the phase transition.  
Could we combine this analysis and that analysis on the
time history of $N_{CS}(t)$ to see the local structure of the universe? 

     The nucleated bubble expands in the medium.  The bubble wall 
is accelerated by the pressure inside the bubble, while decelerated  
by the scattering of thermal particles in the outside medium.  
The velocity of the wall 
$v_{w}(t)$ first grows exponentially and approaches gradually to a
constant value $v_{w}(\infty)$ determined by the balance 
of the different forces.  However, non-equilibrium interactions of the 
thermal plasma with the wall make the calculation of the wall 
velocity rather complicated.  

\medskip
{\bf [Problem 7]} Study the elecroweak phase transition and estimate 
rigorously the various obscure quatities, 
such as $T=T_c-\Delta$, $v_w(t)$, $\delta_w$, and $R_c$.

\medskip 
     After expanding for some time, bubbles fuse with each other, 
which is a very complicate process.  
There exists an exactly solvable model for this  
dynamics, called Kolmogorov-Avrami theory \cite{K-A}. This is a
stochastic theory based on the Poisson distribution of the bubble
formation. However, it is assumed that the  
wall velocity is independent of time, $v_w(t)=$ constant, 
and the critical radius is zero, $R_c=0$. 
A model without such specific assumptions is welcomed.  

\medskip
{\bf [Problem 8]} 
     By modifying the Kolmogorov-Avrami theory, construct a realistic 
theory of the phase transition in field theoretical manner, 
in which the bubbles of three-dimensional domain are created and annihilated. 

\subsection{Other issues}

     Various time scales enter the discussion of baryogenesis.   
The phase transition temperature is roughly given by 
$T\approx 100$ GeV, which is the same order of magnitude as 
the weak boson masses.  The length scale and the time scale 
are respectively given by $\ell_{0} = 2\times 10^{-18}$ m 
and $\tau_{0}= 10^{-26}$ s. 
The time scales of the weak and strong interactions may be estimated 
by second order processes:  
\begin{eqnarray}
 \tau_{weak} &=& \frac{1}{\sigma_{weak}n} \approx \frac{1}{\alpha_w^2 T} 
                                 \approx\frac{10^3}{T},          \\
 \tau_{strong} &=& \frac{1}{\sigma_{strong}n} \approx \frac{1}{\alpha_s^2 T} 
                                 \approx\frac{10^2}{T},          
\end{eqnarray}
where $n$ denotes the number density of the relevant particles 
approximated by $n\approx T^3$.  
The time scale of the baryon number violation is estimated by 
\begin{equation}
 \tau_{N_B-violation}= \frac{n}{\Gamma_{N_B-violation}} 
\approx \frac{1}{\alpha_w^4 T}\approx \frac{10^6}{T}. 
\end{equation}
To see whether the system placed in the experimental apparatus 
of the universe is in thermal equilibrium or not, 
the time scales of interactions have to be compared with that of 
the expansion of the universe  
\begin{equation}
    \tau_{expansion}\approx\frac{1}{H}
 \approx \sqrt{\frac{M_{Planck}^2}{T^4}}\approx \frac{10^{17}}{T},
\end{equation}
where $H$ denotes the Hubble constant. 
Comparing these results, we have the ratio 
\begin{equation}
\tau_{strong}:\tau_{weak}:\tau_{N_B-violation}:\tau_{expansion} = 
                10^{-1} :1 : 10^{3} :10^{14}. 
\end{equation}
Therefore, the effect of the expansion of the universe 
is completely negligible for the weak interaction, 
strong interaction,  and $N_B$-violating transition.  
The $N_B$-violating transition is rather slow behaving as
if the fourth order process of the weak interaction.   

     The amount of baryon asymmetry generated in the phase 
transition depends on the transport time $\tau_{T}$ of 
a net hypercharge emitted from the wall.   
This represents a time within which the particles 
carring the net hypercharge can
travel in the medium without suffering so much from 
thermal interactions with the medium.   
For the leptons and the quarks, $\tau_T$  
may be given by $\tau_{weak}$ and  $\tau_{strong}$, respectively.  
After moving approximately at the speed of light 
during the transport time, the particles get thermalized to 
bias equilibrium conditions for favoring a non-vanishing 
value of baryon number, which is then generated by 
electroweak anomaly.  
The produced baryon number is swallowed from behind 
by the macroscopic bubble wall moving roughly 
at a constant speed $v_w$.  
The time for the wall to catch up with the baryon number is thus given by 
$\tau_{catch up}\simeq\tau_T/v_w$.  
This period is not so large, but should be sufficiently large for 
the system to establish thermal equilibrium and the $N_B$-violating transition
to occur.   

     The net flux of hypercharge is emitted from the bubble wall 
to the symmetric phase, owing to $CP$ violation.  
In the processes that the particles are reflected or transmitted 
at the wall, $CP$ violation makes differences in their probabilities 
between $CP$-conjugate states, which leads to 
the non-vanishing density of hypercharge.  
For obtaining quantitatively the hypercharge flux, it is necessary to 
solve Dirac equations or Klein-Gordon equations.  
The methods of solving these equations numerically 
have been established and work very well now.  
On the other hand, analytical methods have been applied only for  
limited cases.   
 
\medskip
{\bf [Problem 9]} Obtain analyically the $CP$ asymmetry in the reflection and
transmission probabilities for the particles at the wall.  Can we use the
exact solvability in the soliton physics for this purpose? The scattering
data and the potential are strongly related in the soliton theory. 
 
\medskip
{\bf [Problem 10]} Discuss the conservation of angular momentum and 
the mixing of left-handed and right-handed particles, when the particles 
enter the bubble wall obliquely. 

\medskip
     The electroweak baryogenesis necessitates new sources of 
$CP$ violation other than the Kobayashi-Maskawa phase of the SM.  
Up to now several models have been proposed.  If 
the new $CP$-violating source of a model affects phenomena observable in 
experiments,  the model could be constrained nontrivially.  
In these models detailed analyses for consistency between the 
baryon asymmetry and the experimental observables are indispensable.  
On the other hand, in a recently proposed 
model \cite{funakubo} the new $CP$-violating source  
could be only viable within the bubble wall, which does not affect 
phenomena in the broken phase. 

     The microscopic picture for various transitions of the particles 
outside the bubble wall has not been well established.      
The particles reflected or transmitted from the bubble wall 
interact with the medium and then $N_B$-violating transitions occur.   
These processes depend on 
the details of the reactions for the particles.  
Although analyzing such a whole system is very complicated,  
the microscopic picture is necessary for discussing 
the electroweak baryogenesis more quantitatively.  

\section*{Acknowledgments}

     We give our sincere thanks to Kaoru Hagiwara without whom 
this series of KEK meetings had not been organized so 
splendidly and to whom all the participants are indebted.
We are grateful to Hajime Aoki, Jiro Arafune, Koichi Funakubo, 
Jun-ichi Kamoshita, Yasuhiko Katsuki, Minako Kitahara, Takao Ohta,
Isamu Watanabe, Yoshio Yamaguchi, and Katsuji Yamamoto for fruitful and
valuable discussions.  
The work of M.A. is supported in part by 
the Grant-in-Aid for Scientific Research 
from the Ministry of Education, Sciences and Culture, in Japan. 
This work is supported in part by the Grant-in-Aid for Scientific Research
(No. 08640357) and by the Grant-in-Aid for Joint Scientific Research (No.
08044089) from the Ministry of Education, Sciences and Culture, in Japan. 


\end{document}